\documentclass[12pt]{article}
\usepackage{cyr}
\usepackage{url}
\usepackage{epsf}
\usepackage{pazh}

\usepackage[usenames]{color}
\usepackage{colortbl}
\usepackage{color}

\usepackage{natbib}
\usepackage{lscape}
\usepackage{amssymb}
\usepackage{amsmath}
\usepackage[dvips]{graphicx}
\usepackage{array}

\usepackage[utf8]{inputenc}
\usepackage[T2A]{fontenc}
\usepackage[english]{babel}

\tightenlines

\newcommand{\gro}{GRO J2058+42}
\newcommand{\nus}{NuSTAR}
\graphicspath{{pictures/}}
\DeclareGraphicsExtensions{.pdf,.png,.jpg,.eps}
\DeclareUnicodeCharacter{2212}{-}

\usepackage{natbib}
\usepackage{array}
\newcolumntype{H}{>{\setbox0=\hbox\bgroup}c<{\egroup}@{}}

\usepackage[mathscr]{eucal}

\usepackage{multirow}
\begin{document}
\baselineskip 21pt

\title{Changes in the nature of the spectral continuum and the stability of the cyclotron line in the X-ray pulsar  \gro}

\author{\bf  A. S. Gorban\affilmark{1,2*}, S. V. Molkov\affilmark{1}, S.S.Tsygankov\affilmark{3,1}, A.A. Mushtukov\affilmark{4,1} and A. A. Lutovinov\affilmark{1,2}}

\affil{
{\it (1) Space Research Institute, Russian Academy of Sciences, Profsoyuznaya ul. 84/32, Moscow, 117997 Russia}\\
{\it (2) Higher School of Economics, National Research University, Myasnitskaya ul. 20, Moscow, 101100 Russia}\\
{\it (3) Department of Physics and Astronomy, FI-20014 University of Turku, Finland}\\
{\it (4) Leiden Observatory, Leiden University, NL-2300RA, Leiden, The Netherland}
}

\sloppypar
\vspace{2mm}
\noindent
The results of the broadband spectral and timing study of the transient X-ray pulsar \gro\ in a wide energy range at a low luminosity $L_{x} \simeq 2.5\times 10^{36}$ erg s$^{-1}$ are reported.
The data revealed that the pulse profile and pulse fraction of the source are significantly changed in comparison with previous NuSTAR observations, when the source was ten times brighter. The cyclotron absorption line at $\sim10$ keV in the narrow phase interval is consistent with the high state observations. 
Spectral analysis showed that at high luminosities $L_{x}\simeq (2.7-3.2)\times 10^{37}$ erg s$^{-1}$ the spectrum has a shape typical of accreting pulsars, while when the luminosity drops by about an order of magnitude, to $2.5\times 10^{36}$ erg s$^{-1}$ a two-component model is necessary to its describing. This behavior fits into a model in which the low-energy part of the spectrum is formed in a hot spot, and the high-energy part is formed as a result of resonant Compton scattering by incident matter in an accretion channel above the surface of a neutron star.

\vfill
\noindent\rule{8cm}{1pt}\\
{$^*$ E-mail: $<$gorban@iki.rssi.ru$>$}

\clearpage
\section{INTRODUCTION}

The transient X-ray pulsar \gro\ was discovered in September 1995 during its giant outburst with the BATSE instrument BATSE on board the {\it Compton Gamma Ray Observatory}  (Wilson et al., 1995). The type II outburst (bright outbursts in Be systems independent of the binary orbital phase with a peak luminosity reaching the Eddington limit for a neutron star) lasted about 46 days with the maximum flux about 300 mCrab in the 20-100 keV energy range. The pulse period is decreased from 198 s to 196 s during these days.   
After the main burst \gro\ had remained active for approximately two years. The pulsar demonstrated a series of weaker type I bursts (weak bursts associated with the passage of a neutron star through the periastron of a binary system) with the maximum pulsed flux of 10-15 mCrab (20-50 keV). The intervals between bursts was about 110 days, which was interpreted as an orbital period in the binary system. Small increases in brightness up to $\sim$8 mCrab were also observed between the main flares (Wilson et al. 1995). These outbursts were confirmed by the RXTE/ASM instrument and indicated a possible shorter orbital period of 55 days (Corbet et al. 1997; Wilson et al. 2005).

After the period of activity \gro\ went into quiescence and exhibited no outburst activity in X-rays until March 22, 2019, when the Burst Alert Telescope (BAT) onboard the {\it Neil Gehrels Swift} observatory (Gehrels et al., 2004) and the Gamma-ray Burst Monitor (GBM) onboard the {\it Fermi}  observatory (Meagan et al. 2009) detected a new giant outburst (Bartelmy et al. 2019; Malakaria et al. 2019). The peak of luminosity this outburst was similar to the previous outburst. During the 2019 outburst the source was monitored by {\it Swift, NICER, Fermi} observatories. Two observations were performed also with the {\it NuSTAR} observatory near the maximum luminosity (Molkov et al. 2019).

The periodic bursts, which follow the giant outburst, point that the optical companion of binary system is a Be star. The optical companion was accurately determined with the optical photometry and spectroscopy (Reig et al., 2005) and was identified with the star of the O9.5-B0IV-Ve type, which is located at a distance of $9 \pm 1$ kpc.
The distance to the binary system was estimated as ${8.0_{-1.2}^{+0.9}}$ kpc from the Gaia telescope data 
(Arnason et al. 2021). This value will be used later in the article.

Molkov et al. (2019) discovered a cyclotron resonant scattering line at $\sim10$ keV in the source spectrum and its harmonics at $\sim20$ and $\sim30$ keV by using data from the {\it NuSTAR} observatory. It was shown that the cyclotron line and its harmonics were registered only in a narrow range of pulsar rotation phases. Based on these measurements the magnetic strength of the pulsar was estimated as $\sim10^{12}$ G.

An analysis of the {\it AstroSat} observatory data confirmed the presence of the cyclotron line in the \gro\ spectrum  (Mukerjee et al., 2020) and significant variations its parameters with the pulse phase. Moreover, authors reported about a discovery of quasi-periodic oscillations with the frequency of 0.09 Hz in the source power spectrum.

In order to study the properties of the pulsar at low accretion rates (see, for example, the review by Tsygankov et al. 2017 of the properties of pulsars in Be systems in low states), we asked another observation with the {\it NuSTAR} observatory, which was carried out approximately 150 days after the maximum of the main outburst. The flux $F_3=(3.3_{-2.8}^{+0.4})\times 10^{-10}$ erg cm$^{-2}$ s$^{-1}$ was measured from the source in the energy range of 3-79 keV. This value is about an order of magnitude lower than the fluxes detected from the source during the first two {\it NuSTAR} observations: $F_1=(3.6\pm0.1)\times 10^{-9}$ erg cm$^{-2}$ s$^{-1}$ and $F_2=(4.3\pm{0.1)\times 10^{-9}}$ erg cm$^{-2}$ s$^{-1}$ (Molkov et al. 2019). 

This paper is devoted to studying the characteristics of a pulsar in the low state and comparing them with those previously obtained for the bright state. 

\section{OBSERVATIONS AND DATA ANALYSIS}

\begin{table}
\caption{Observations of {\gro}, used in this paper}
\centering
\label{tab:data}
\begin{tabular}{cccc}	\\
\hline 
Telescope & Date & ObsID & Exposure \\ [0.5ex] 
\hline
NuSTAR&25.03.2019 (MJD 58567)&90501313002&20 ksec \\
NuSTAR&28.08.2019 (MJD 58723)&90501336002&38 ksec \\
Swift/XRT&28.08.2019 (MJD 58723)&00088982001&2 ksec \\ [1ex]
\hline
\end{tabular}
\label{table:nonlin}
\end{table}

The NuSTAR observation of source with low state luminosity was carried out on August 28, 2019 (MJD 58723) with an exposure time 38 ksec (ObsID 90501336002). 
The observatory \nus\ (Nuclear Spectroscopic Telescope ARray) is sensitive to X-rays in the energy range of 3 - 79 keV. It consists of two co-aligned identical telescopes (FPMA and FPMB) with an energy resolution of $\sim400$ eV at 10 keV (Harrison et al., 2013)

The \nus\ data were analyzed with the standard software (NuSTARDAS v1.9.7), which is supplied as part of the HEASOFT v6.29 software with CALDB v20211202 calibrations.

Firstly, data processing was carried out with the NUPIPELINE procedure. Further, we used NUPRODUCTS to obtain energy spectra and light curves of the source. The \gro\ data of both modules were extracted from a circular region of radius 50'' centred on the source position. The background was extracted with a similar radius of 50'' on the same detector chip. To compare properties of the pulsar in different intensity states, we used {\it NuSTAR} data from one of the previous observation, which was carried out at March 25, 2019  (ObsID 90501313002) with the exposure of $\sim20$ ksec (see Table \ref{tab:data}). The background was extracted on the adjacent chip due to a high brightness of the source. Photons from the source and background were chosen from a circular region of radius 70''.

\begin{figure}
	\center{\includegraphics[scale=0.65]{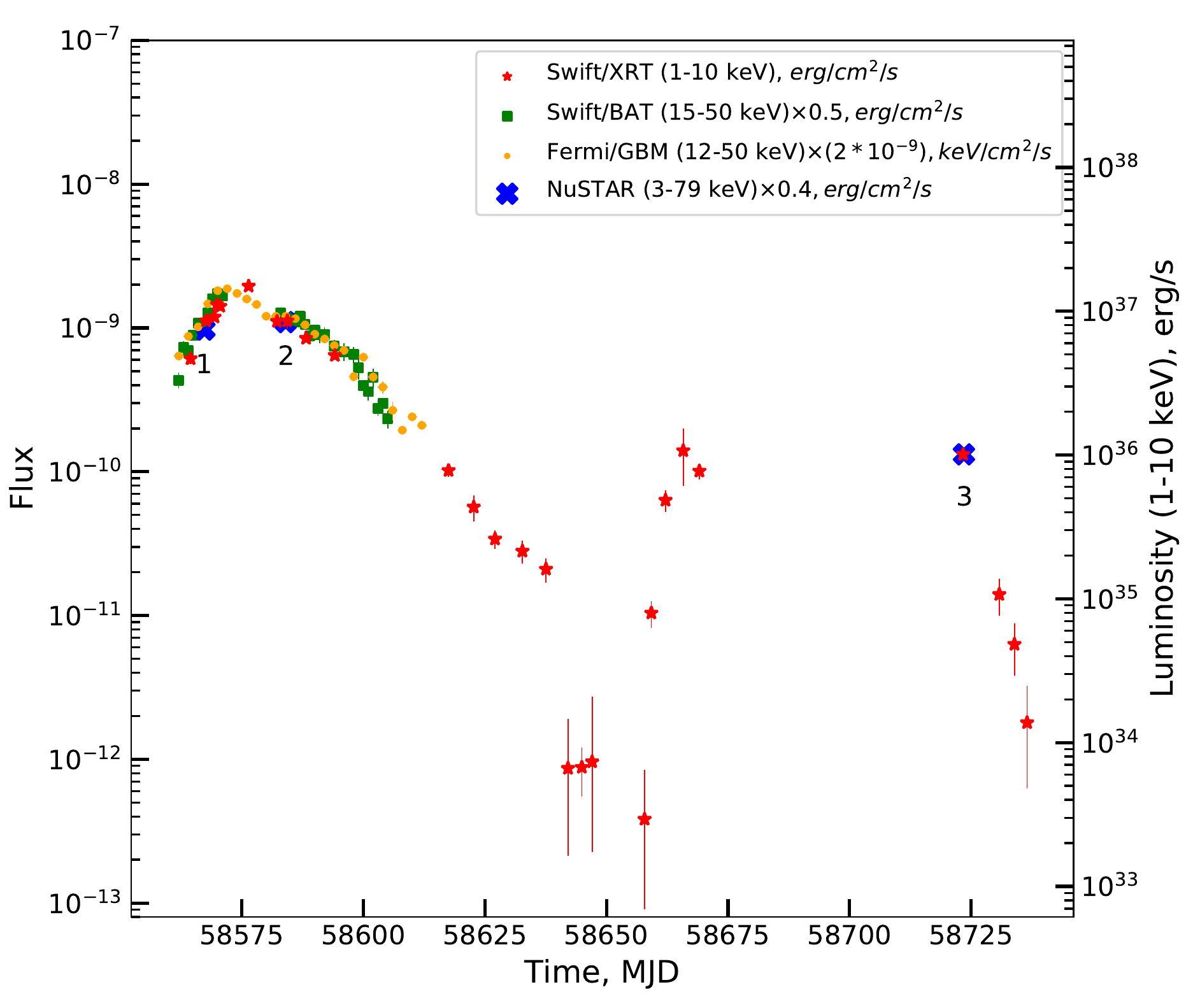}}
	\caption{The light curve of X-ray pulsar \gro\ in several energy ranges obtained with different instruments. The fluxes are given in different units of measurement and re-normalized for the better clarity (see the inset). Numbers point to the NuSTAR observations. Luminosity is shown in the energy range of 1-10 keV to visualize the light curve profile in this picture.}
	\label{fig:lc}
\end{figure}

To expand our spectral and timing analysis we used a Swift/XRT observation with the exposure time of 2 ksec (ObsID 00088982001), which was made simultaneously with the \nus\ observation.  The telescope Swift/XRT works in the energy range of 0.3-10 keV. The source spectra were obtained with the online tools (Evans et al. 2009) provided by the UK Swift Science Data Centre at the University of Leicester.\footnote{\url{http://www.swift.ac.uk/user\_objects/}}
We used the Photon Counting (PC) mode data for the analysis. The energy spectra were also binned by 25 counts per bin for spectral analysis with $\chi^2$ statistic. The approximation of the spectra was carried out in the {\sc XSPEC} v12.12.0  package (Arnaud et al., 1999).


\section{RESULTS}

Several \gro\ observations made by the NuSTAR observatory during its outburst activity in 2019 allowed us to make a detailed comparative analysis of the temporal and spectral properties of the pulsar in states with an accretion rate that differs by about order of magnitude: bright state luminosity is $L_{x}\simeq 2.7\times 10^{37}$ erg s$^{-1}$ (ObsID 90501313002), low state is $L_{x} \simeq 2.5\times 10^{36}$ erg s$^{-1}$ (ObsID 90501336002) (here and further, value for energy range 3-79 keV uses). The light curve obtained from the different instruments is shown in Fig.1. 

\begin{figure}
	\center{\includegraphics[scale=0.5]{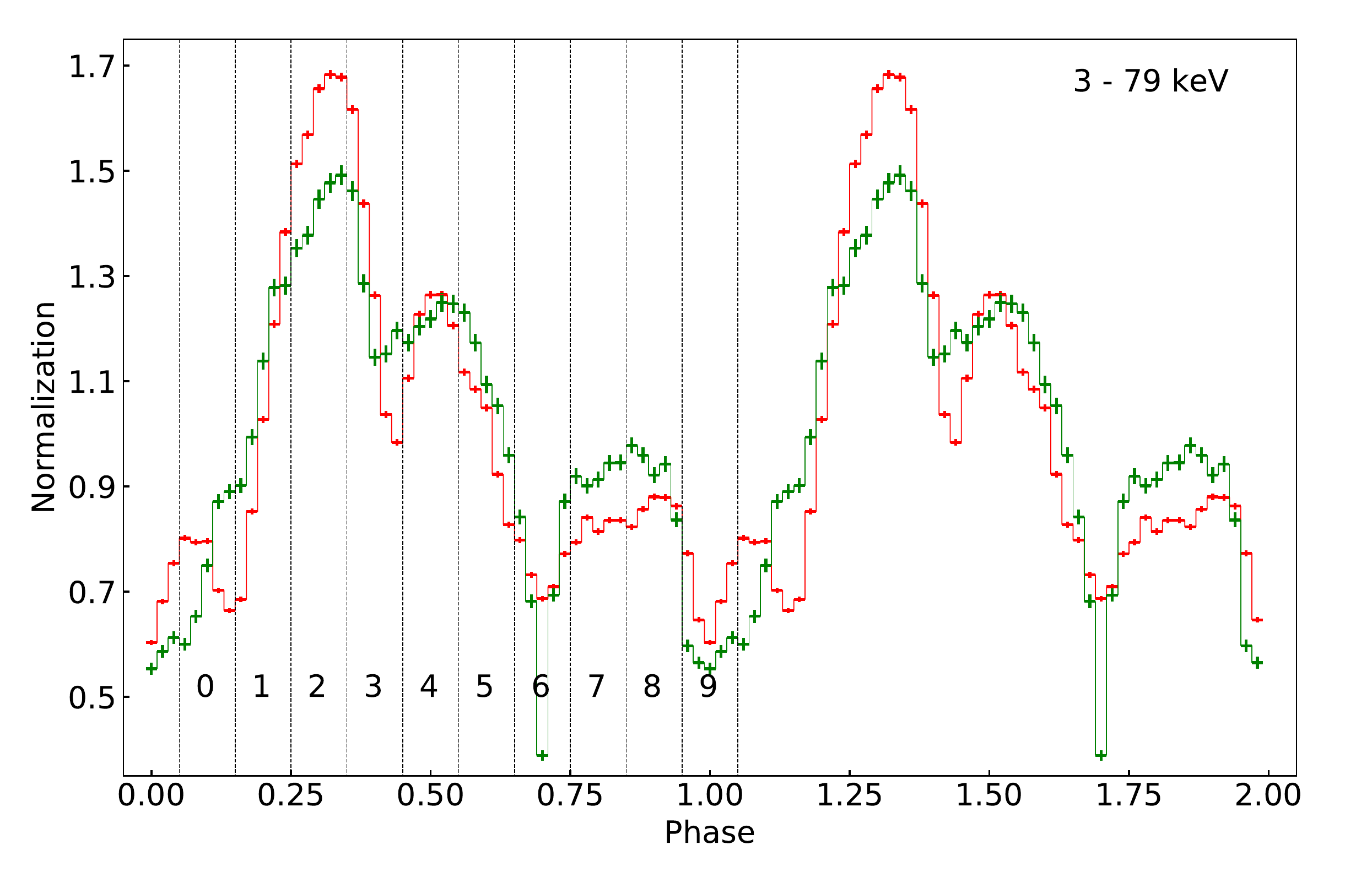}}
	\caption{Pulse profiles \gro\ in wide energy range 3-79 keV from NuSTAR observatory. Red lines show bright state observation in March 25, 2019 (ObsID: 90501313002), green lines - low state observation in August 28, 2019 (ObsID: 90501336002).} 
	\label{fig:lc3_79}
\end{figure}

\subsection{Timing Analysis of the Emission from \gro}

For timing analysis of the {\it NuSTAR} data, photon arrival times were first corrected for the barycenter of the Solar System using standard HEASOFT package tools.
Due to the lack of well-measured orbital parameters of the system, the photon arrival time was not corrected for the orbital motion of the neutron star. All light curves were obtained with a time resolution of 0.1 s, then the background was subtracted from them, and the light curves of the two modules were combined using the lcmath tool (FTOOLS V6.29). Further, we determined the pulse period of 193.375$\pm$0.001s using the epoch folding technique (program efsearch from the HEASOFT package). The error in the period was estimated using Monte Carlo simulations of the light curve (see Boldin et al. 2013 for details)
\begin{figure}
	\center{\includegraphics[scale=0.7]{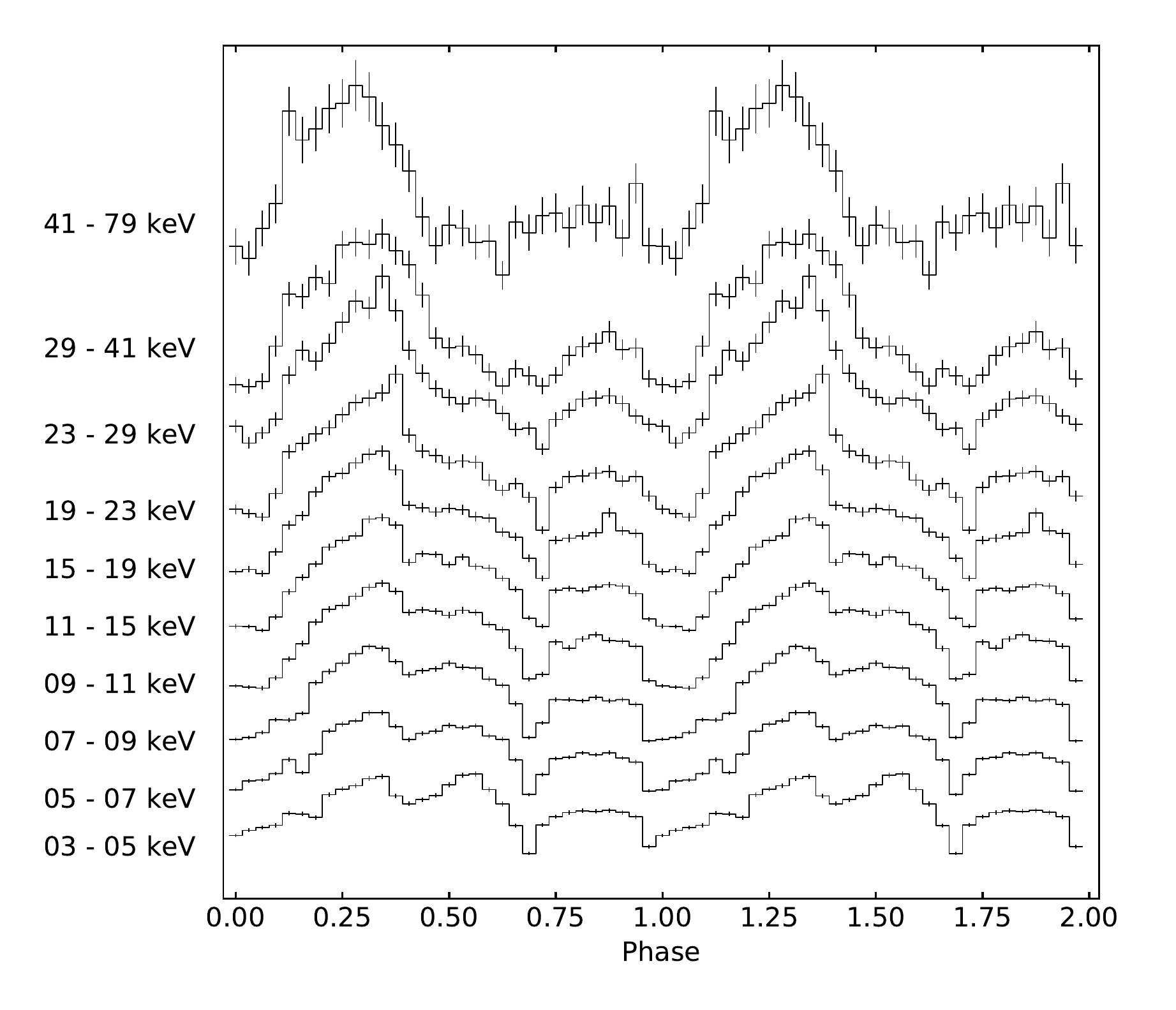}}
	\caption{Pulse profiles of \gro\ in low state in ten energy bands from the NuSTAR data in August 2019. Profiles are shifted along the Y axis for clarity.} 
	\label{fig:all}
\end{figure}

The measured pulse period was used to obtain the pulse profile in a wide energy range of 3-79 keV and comparison with the pulse profile in the bright state (Fig. \ref{fig:lc3_79}). It is clearly see that pulse profile was changed between two observations. The main differences in the profiles are expressed in the main peak -- in the bright state at phases 0.25-0.35 its relative intensity is much higher than in the low state. At the same time, a clearly visible deep narrow minimum appears in the low state at phases 0.65-0.75. Also, an additional peak is observed in the bright state at phases of 0.1, which possibly associated with a more complex structure of emission regions on the surface of a neutron star.

\begin{figure}
	\center{\includegraphics[scale=0.6]{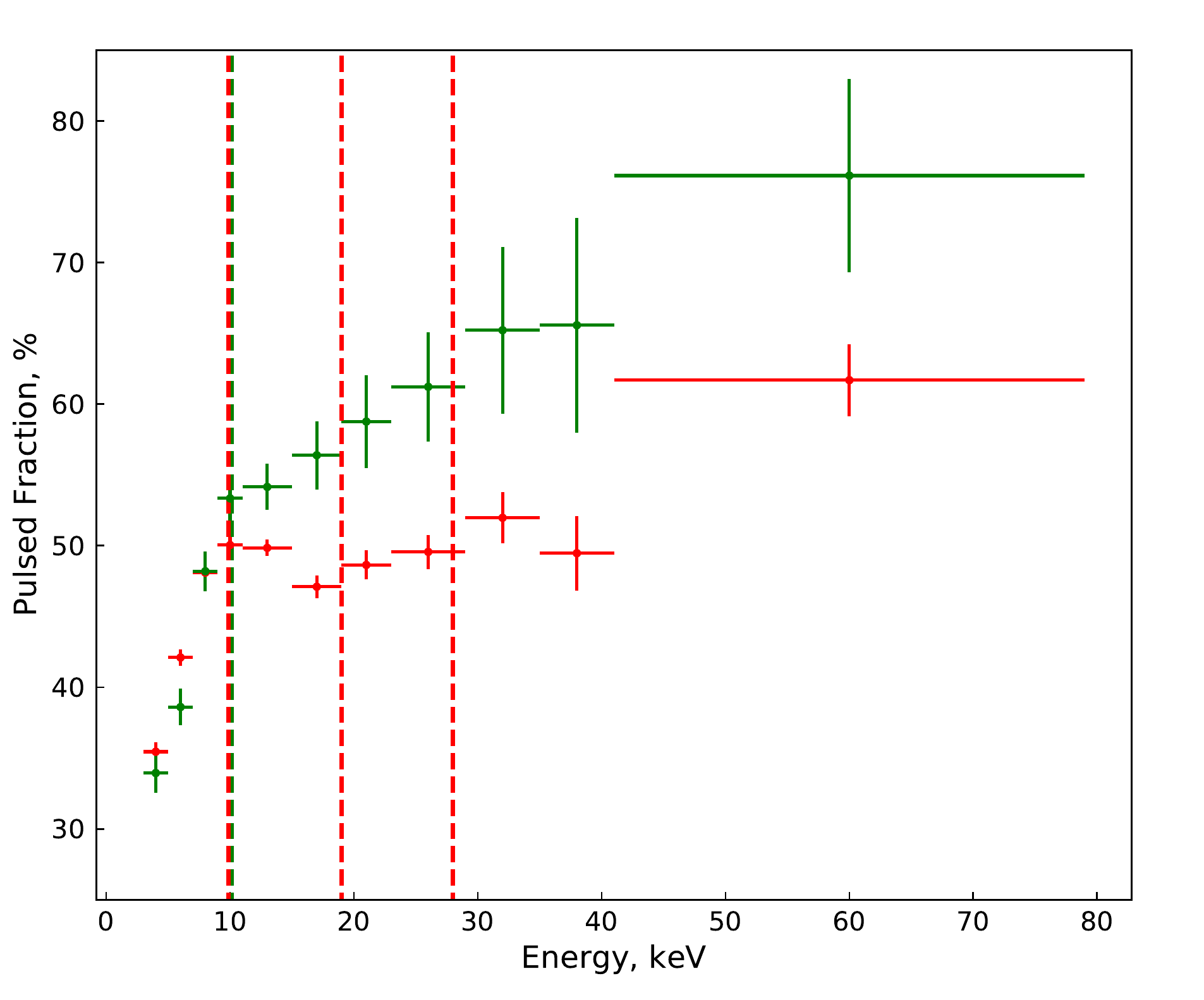}}
	\caption{Pulsed fraction for \gro\ as a function of energy for two NuSTAR observations: red crosses correspond to the  bright state (March 25, 2019; ObsID: 90501313002), green color --  low state (August 28, 2019; ObsID: 90501336002). Dashed lines indicate positions of the cyclotron line and its harmonics.} 
	\label{fig:pf}
\end{figure}

At the next step we extracted light curves in the energy ranges of 3-5, 5-7, 7-9, 9-11, 11-15, 15-19, 19-23, 23-29, 29-41 and 41-79 keV and constructed corresponding pulse profiles (Fig. \ref{fig:all}). The figure demonstrates that the pulse profile is changed significantly with the energy. Particularly, in the energy range from 3 keV to $\sim$10 keV, the profile shows four separate peaks at phases of 0.1, 0.3, 0.6, as in the bright state (see Molkov et al. 2019). Also, a broad peak near phases 0.7-0.9 is appeared in the low state. As the energy increases, the peak at phase 0.3 expands as in the bright state, the peak at phase 0.7-0.9 persists at all energies up to $\sim40$ keV. At higher energies, the secondary peak disappears, which leads to the appearance of a relatively flat region in a wide phase range of 0.5–1.0. This shape of the pulse profile was also observed previously at energies of 20–70 keV (Wilson et al. 1998). 

We have plotted the dependence of the pulse fraction (PF) on the energy for both \nus observations. The PF was defined as the ratio (Fmax$-$Fmin)/(Fmax+Fmin), where Fmax and Fmin are the maximum and minimum fluxes in the pulse profile consisting of 20 bins, respectively. According to Fig.\,\ref{fig:pf}, the PF in both states increases with the energy, which is typical of most X-ray pulsars (Lutovinov and Tsygankov 2009). However, this growth is non-monotonous. In particular, in the bright state (red dots), some feature (local increase of the PF) is clearly visible around 10 keV, where the cyclotron absorption line is located. In a low state (blue dots), a broken dependence is also shown around  energy. Peculiarities at energies corresponding to the higher harmonics of the line are not visible due to insufficient statistics and the large width of the energy channels. It is important to note that the presence of similar features in the PF near the cyclotron energy was shown earlier for a number of X-ray pulsars (Tsygankov et al. 2007; Lutovinov and Tsygankov 2009; Tsygankov et al. 2010; Lutovinov et al. 2017).


\subsection{Spectral Analysis of the Source \gro\ }

We used continuum model in the form of thermal Comptonization ({\sc compTT} in the XSPEC package, Titarchuk, 1994). This model was applied by Molkov et al. (2019) for description of bright state spectra, that allow to compare the spectral properties of \gro\ in different intensity states. In addition, the iron emission line at $\sim6.4$ keV was added to the model. Its width was frozen at 0.24 keV determined by Molkov et al. (2019).  

The interstellar absorption was taken into account by adding the component {\sc phabs} into the model. An obtained value of absorption $N_{\rm H} \simeq 7\times 10^{21}$ cm$^{-2}$ is in good agreement with the Galactic absorption in the direction to the source $\sim 6.2\times 10^{21}$ cm$^{-2}$ (HI4PI Collaboration, N. Ben Bekhti, L. Floer, et al., 2016).
In order to take into account the different calibrations of the FPMA and FMPB modules aboard {\it NuSTAR}, as well as the nonsimultaneity of the observations of {\it NuSTAR} and {\it Swift}, normalization coefficients were added as well (see Table \ref{tbl_2}). The rest of the model parameters for different instruments were fixed among themselves.

\begin{figure}[ht]
	\center{\includegraphics[scale=0.75]{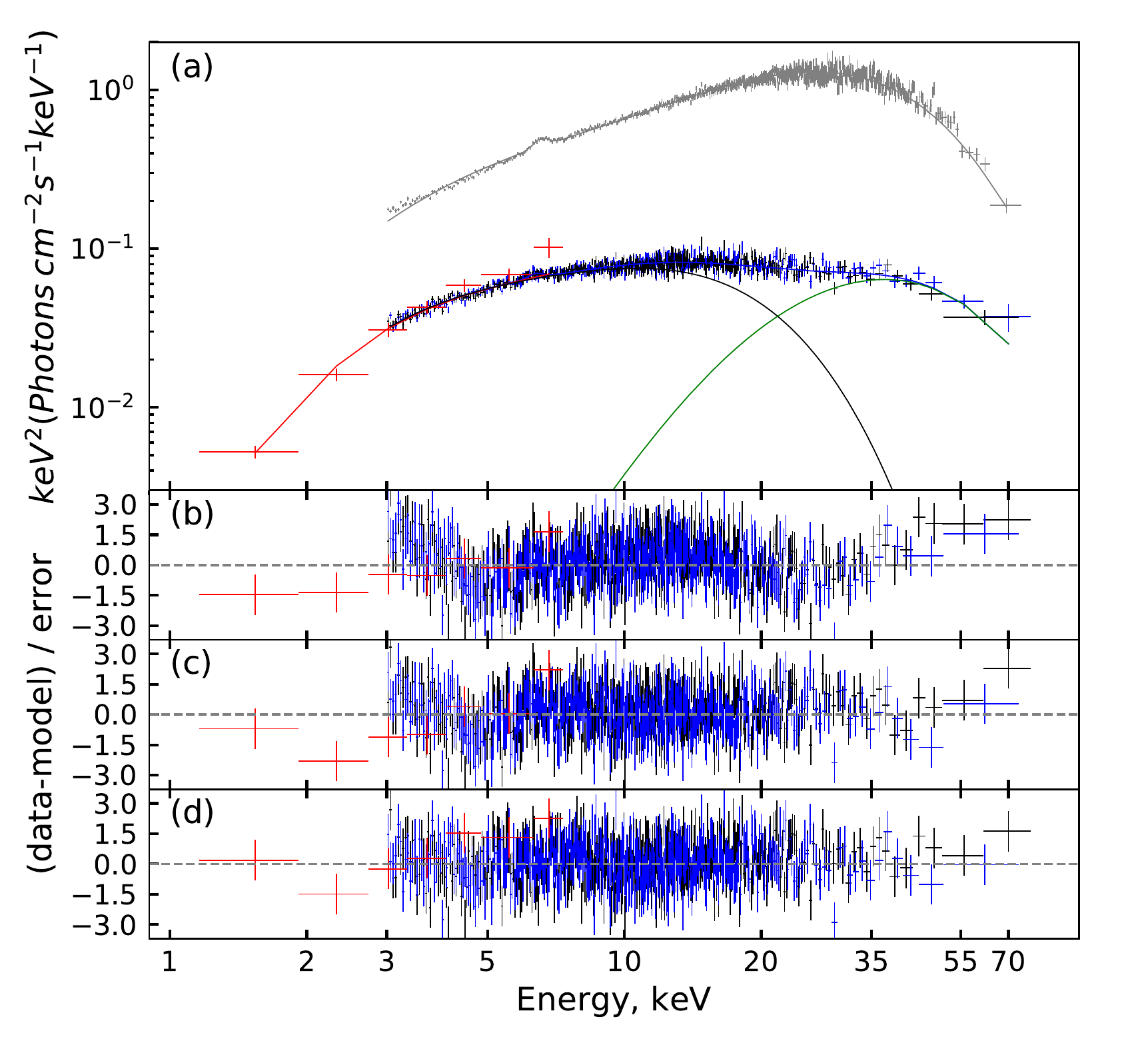}}
	\caption{(a) Energy spectrum \gro\ measured in August 2019 by {\it NuSTAR} (ObsID 90501336002) (green and black dots) and {\it Swift}/XRT (red dots), the solid lines show the best-fit model; gray dots show the energy spectrum obtained by {\nus} in March 2019 (ObsID 90501313002). Panel (b) shows the residuals of the observational data from the {\sc phabs*(gaussian+compTT)} model, panel (c) is for the {\sc phabs*(gaussian+compTT)*gabs} model, panel (d) shows the residuals of the observed data from the {\sc phabs*(gaussian+compTT+compTT)} model.} 
	\label{fig:spec}
\end{figure}

In contrast to the data obtained in the bright state, for the low state, the {\sc phabs*(gaussian+compTT)} model shows an unsatisfactory quality of the fit  with $\chi^{2}$ = 1443.92 for 1218 degrees of freedom (d.o.f.) for the \gro\ broadband spectrum. Also, there are noticeable deviations in around the energy of $\sim$27 keV (Fig. \ref{fig:spec}, b). Formally, to describe this feature, we added a wide absorption line with a Gaussian profile {\sc gabs} to the model.
This resulted in a significant improvement in the quality of the fit with $\chi^{2}$ = 1262.05 (1216 d.o.f.) (Fig. \ref{fig:spec}, c). However, a sufficiently large width of the line (see Table \ref{tbl_2}) and the cyclotron absorption lines discovered earlier by Molkov et al. (2019) at other energies cast doubt on the physical meaning of this component. Therefore, we took a different approach to describe the spectrum and used a combination of two Comptonization components (the {\sc compTT + compTT} model in XSPEC) as a continuum. The temperatures of the seed photons of both components were tied to follow the approach of Tsygankov et al. (2019). This model {\sc phabs*(gaussian+compTT+compTT)} showed best quality fit with $\chi^{2}$ = 1219.86 for 1214 d.o.f. (fig. \ref{fig:spec}, d). The model parameters are presented in Table \ref{tbl_2}.

\begin{figure}[th]
	\center{\includegraphics[scale=0.75]{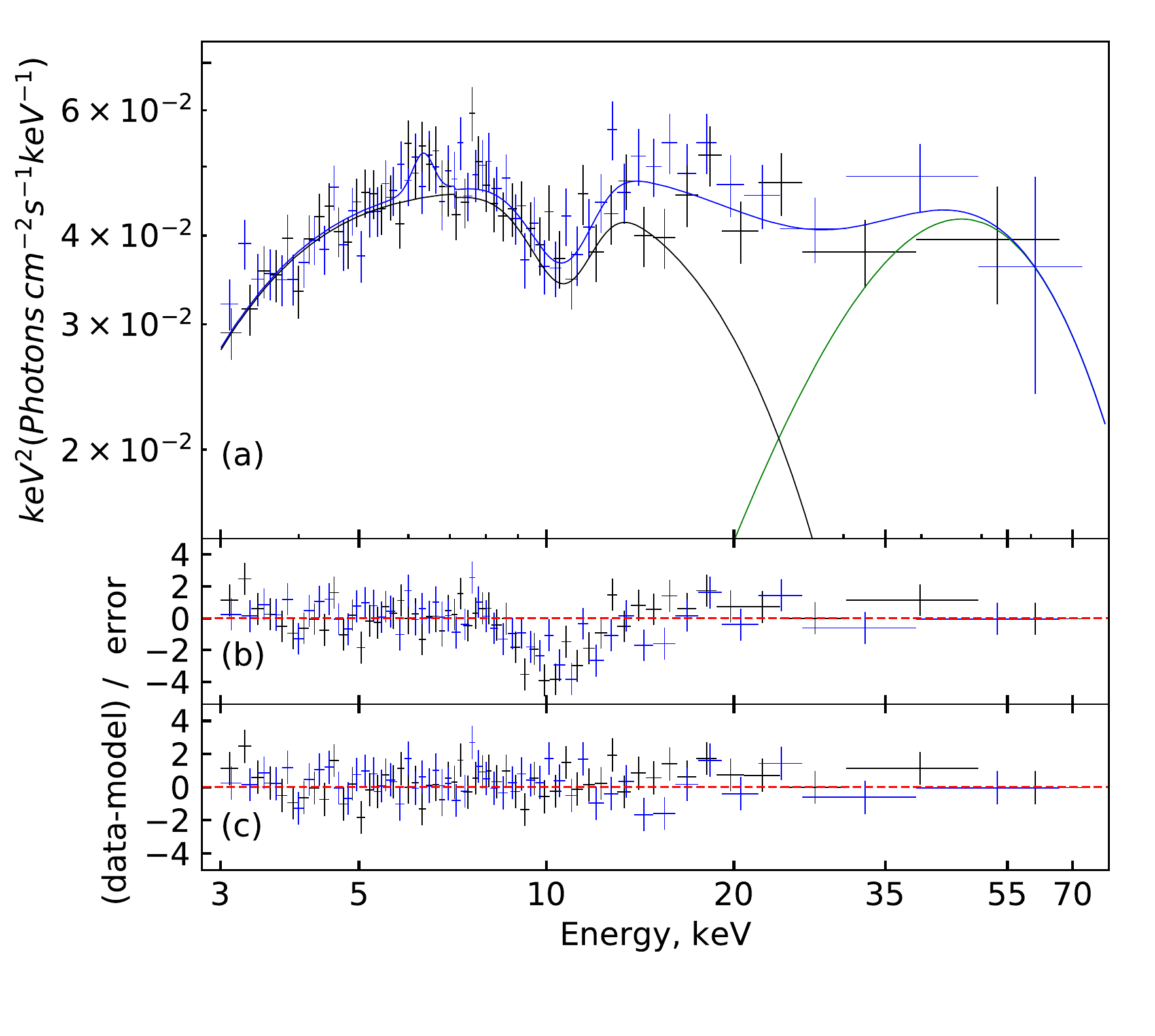}}
	\caption{ (a) Energy spectrum of GRO J2058+42 at the pulse phases 0.05-0.15 for NuSTAR observation in August 2019 (ObsID 90501336002). Both FPMA and FPMB data are shown as blue and black dots. The solid blue line shows the best-fit model, and the black and green lines show the contribution of different components. Panel (b) shows residuals of the observational data from the {\sc phabs*(gaussian+comptt+comptt)} model similar to the average spectrum model, (c)  for the model with the cyclotron line {\sc phabs*(gaussian+comptt+comptt)* gabs}.} 
	\label{fig:phase1}
\end{figure}

Previously, Molkov et al. (2019) showed that the phase-resolved spectra in the bright state contain a cyclotron line at $\sim$10 keV at pulse phases 0.05-0.15, as well as its harmonics at $\sim$20 and $\sim$30 keV. We also carried out phase-resolved spectroscopy of the third NuSTAR observation.
We used model {\sc phabs*(gaussian+compTT+compTT)} for phase-resolved spectra, just as for the averaged spectrum. 
This model without additional components showed the adequate description of all phased spectra except for phases 0.05-0.15. The fit of the spectrum at these phases revealed unsatisfactory quality with $\chi^{2}$ = 495.61 for 406 d.o.f. due to the feature around 10 keV (Fig. \ref{fig:phase1}, b). To describe this feature we added a gabs component of about 10 keV (fig. \ref{fig:phase1}, c), that significantly improved quality of fit with $\chi^{2}$ = 451.09 for 403 d.o.f. (see Table \ref{tbl_3}). Thus, the cyclotron absorption line at $\sim10$ keV was confirmed in the source spectrum at the low state at the same phases as in previous observations. We were unable to register harmonics at energies of 20 and 30 keV due to the lack statistics in the phase-resolved spectrum.

\begin{table}
\caption{Parameters of the spectrum for \gro\ with the continuum described by the {\sc compTT+gabs} and {\sc compTT+compTT}}
\centering
\label{tbl_2}
\begin{tabular}{ccc}	\\
\hline 
Model parameters & {\sc compTT+gabs} & {\sc compTT+compTT}  \\ [0.5ex] 
\hline
$N_{H}$, {$10^{22}$ cm$^{-2}$}&${0.2\pm{0.1}}$&${0.7_{-0.2}^{+0.2}}$ \\
$E_{Fe}$, keV&6.4, frozen&$6.32_{-0.08}^{+0.08}$ \\
$\sigma_{Fe}$, keV&0.24&0.24, frozen \\                   
$W_{Fe}$, keV&$0.03_{-0.01}^{+0.01}$&$0.04_{-0.01}^{+0.02}$ \\
$E_{Cycl}$, keV& $27.2\pm{0.7}$&$-$ \\                   
$\sigma_{Cycl}$, keV&$10\pm{2}$&$-$ \\
$\tau_{Cycl}$&$0.53\pm{0.15}$&$-$\\
$T_{0, Comptt}, low$, keV&$1.27\pm{0.03}$&$1.01_{-0.06}^{+0.05}$\\
$kT_{Comptt}, low$, keV&$10.9\pm{0.7}$&$3.9_{-0.4}^{+0.4}$ \\
$\tau_{Comptt}, low$&$3.7\pm{0.4}$&$8.0_{-1.8}^{+16.8}$ \\
$T_{0, Comptt}, high$, keV&$-$&= $T_{0, Comptt}, low$ \\
$kT_{Comptt}, high$, keV&$-$&$14.6_{-1.2}^{+1.1}$ \\
$\tau_{Comptt}, high$&$-$&$2.9_{-0.5}^{+0.9}$ \\
$Flux$ (3 - 79 keV), $10^{-10}$&$3.2_{-0.1}^{+0.1}$&$3.3_{-2.8}^{+0.4}$ \\
erg\; cm$^{-2}$ $s^{-1}$& & \\
$C_{NuSTAR}$&$0.99_{-0.01}^{+0.01}$&$0.99_{-0.01}^{+0.01}$ \\
$C_{XRT}$&$0.89_{-0.04}^{+0.04}$&$0.89_{-0.04}^{+0.04}$ \\
\hline
$\chi^{2}$ (d.o.f.) & 1262.05 (1216)& 1219.86 (1214) \\
\hline
&
\end{tabular}
\end{table}

\begin{table}
\caption{Parameters of the phased spectrum at 0.05-0.15 for \gro\ with the continuum described by the {\sc constanta*phabs*(gaussian+compTT+compTT)*gabs} model}
\centering
\label{tbl_3}
\begin{tabular}{cc}	\\
\hline
Model parameters & Value \\ [0.5ex] 
\hline
$N_{H}$, {$10^{22}$ cm$^{-2}$} & ${0.7},$ fixed \\                   
$E_{Fe}$, keV &	$6.32$, fixed  \\		
$\sigma_{Fe}$, keV & $0.24$, fixed \\              
$W_{Fe}$, keV & $0.08_{-0.04}^{+0.04}$ \\
$E_{Cycl}$, keV & $10.6_{-0.2}^{+0.2}$ \\                  
$\sigma_{Cycl}$, keV & $1.3_{-0.2}^{+0.3}$ \\                  
$\tau_{Cycl}$ & $0.31_{-0.02}^{+0.03}$ \\
$T_{0, Comptt}, low$, keV & $1.01,$ fixed \\
$kT_{Comptt}, low$, keV & $4.9_{-0.5}^{+0.7}$ \\
$\tau_{Comptt}, low$ & $ 4.8_{-0.5}^{+0.6}$ \\
$T_{0, Comptt}, high$, keV & = $T_{0, Comptt}, low$ \\
$kT_{Comptt}, high$, keV & $14.5_{-1.6}^{+2.3}$ \\ 
$\tau_{Comptt}, high$ & >15 \\
$C_{NuSTAR}$ & $0.98_{-0.02}^{+0.02}$ \\
\hline
$\chi^{2}$ (d.o.f.) & 336.49 (362) \\
\hline
\end{tabular}
\end{table}



\section{DISCUSSION}

In the present work, it was shown that when the luminosity of the X-ray pulsar \gro\ decreased to a few by $10^{36}$ erg s$^{-1}$, the energy spectrum of the source changed significantly. In particular, the approximation by the thermal Comptonization model, which works well for the bright state of a pulsar, no longer describes the spectrum. This happened in connection with a clear change in the shape of the spectrum, expressed in its flattening in the region of 10-30 keV. As shown above, an acceptable approximation of the spectrum can be obtained by modifying the model by adding either an absorption line at an energy of about 27 keV, or by adding a second emission component with a high temperature. However, the too large width of the absorption feature and the presence of true cyclotron lines in the phase-resolved spectra of the source at other energies exclude the interpretation of spectrum flattening through the cyclotron absorption. 

The appearing of second emission component in the high energy of low luminosity pulsars is a well-established fact (Tsygankov et al., 2019 a, b; Lutovinov et al., 2021; Doroshenko et al., 2021). This behavior is associated with the Comptonization of cyclotron photons in the overheated atmosphere of a neutron star (Mushtukov et al., 2021; Sokolova-Lapa et al., 2021).
In the case of \gro, the cyclotron line is at the lowest energy ($\sim10$ keV) to compare with all sources where such a change in the spectrum was observed.
The high-energy component of the spectrum cannot be associated with the emission and following Comptonization of cyclotron photons in the hot atmosphere of a neutron star due to the fact that the cyclotron line in the \gro\ spectrum is observed at an energy of $\sim 10$ keV.

On the other hand, the relatively low cyclotron energy makes possible and even necessary another mechanism for the formation of a high-energy component in the spectrum. Namely, in the subcritical accretion regime, which we assume for \gro\ at the considered luminosity, the accretion channel has an optical thickness of $<1$ with respect to Thomson scattering. However, at resonance, where the scattering cross section is much larger than Thomson scattering, the accretion channel is optically thick. 

The position of the resonance in the reference frame of the neutron star surface is redshifted due to the Doppler effect.
The red shift changes depending on the direction of photon escape from the accretion channel, and resonant scattering is expected to cover a wide spectral band from $\sim 6$ to $\sim 10$ keV.
Photons that even once experienced resonant scattering by the incident matter, whose velocity is $\sim 0.5c$, leave the accretion channel with an energy much higher than before scattering, which forms the high-energy part of the spectrum. 
Such a mechanism of high-energy part formation of the spectrum implies a different beam pattern of the outgoing radiation of the low-energy and high-energy parts of the spectrum. As the high-energy part of the spectrum is the result of scattering in the accretion channel, it has to have a beam pattern close to fan-shaped. Probably, the observed evolution of the pulse profile with the photon energy is associated with this difference in the beam patterns at high and low energies (see Fig.\,\ref{fig:all}) (Mushtukov et al., 2018).

\section{CONCLUSIONS}

In this paper we presented the results of timing and spectral analysis of the X-ray pulsar \gro\ in a wide energy range based on the data obtained by {\it NuSTAR} and {\it Swift} observatories in August 2019. According to the results, the source was in the low luminosity state with $L_{x} \simeq 2.5\times 10^{36}$ erg s$^{-1}$ which is an order of magnitude lower than previously observed. Timing analysis has shown the changes in the energy-averaged pulse profile in comparison with the bright state. The PF gradually increases with the energy, that is typical of X-ray pulsars. Additionally, the PF indicates the presence of a feature around the cyclotron energy in both low and bright states.

It has been also shown that the energy spectrum has changed greatly upon transition to a low accretion rate. To describe it, we proposed a continuum model consisting of two Comptonization components. The observed shape of the spectrum in the low state can be interpreted in terms of a model where the low-energy part of the spectrum is formed in a hot spot, while the high-energy part is formed as a result of resonant Compton scattering of photons in the range of $\sim6-10$ keV by the incident matter in the accretion channel above the surface of a neutron star. 

The phase-resolved spectroscopy also showed a photon deficit around 10 keV in the pulse phase interval 0.05-0.15. This feature can be interpreted as a cyclotron absorption line. The position of the cyclotron line at $\sim$10 keV in a narrow phase region is consistent with results of observations in the high state.

\section{ACKNOWLEDGMENTS}

This study was carried out using the data obtained with NuSTAR, a Caltech project, funded by NASA and operated by NASA/JPL and the data provided by the UK Swift Science Data Centre (XRT data analysis). In this study we also used the software provided by the High-Energy Astrophysics Science Archive Research Center (HEASARC), which is a service of the NASA/GSFC Astrophysics Science Division. This work was supported by RSF grant no. 19-12-00423.






\begin{references}
\reference{\it {Arnason}, R. M., {Papei}, H.,  {Barmby}, P., {Bahramian}, A., {Gorski}, M.D., MNRAS, 502, 4, 5455-5470 (2021)}\phantom{}{}
\reference{\it {Arnaud}, K., {Dorman}, B.,  {Gordon}, C., Astrophysics Source Code Library, 10005 (1999)}\phantom{}{}

\reference{\it Barthelmy, S. D., Evans, P. A., Gropp, J. D., et al., GRB Coordinates Network, 23985, 1 (2019)}\phantom{}{}
\reference{\it HI4PI Collaboration: Ben Bekhti, N., Floer, L., et al., Astronomy \& Astrophysics, 594, A116 (2016)}\phantom{}{}
\reference{\it {Boldin}, P. A., {Tsygankov}, S. S.,  {Lutovinov}, A. A., Astron. Lett.,\, 39,\, 375 (2013) }\phantom{}{}
\reference{\it Gehrels, N., Chincarini, G., Giommi, P., et al., The Astrophysical Journal, 611, 1005 (2004)}\phantom{}{}
\reference{\it Doroshenko, V., Santangelo, A., Tsygankov, S. S., and Ji, L., Astronomy \& Astrophysics, 647, A165 (2021)}\phantom{}{}
\reference{\it Corbet, R., Peele, A., and Remillard, R., International Astronomical Union Circular, 6556, 3 (1997)}\phantom{}{}
\reference{\it Lutovinov, A., Tsygankov, S., Molkov, S., et al., The Astrophysical Journal, 912, 17 (2021)}\phantom{}{}

\reference{\it Lutovinov, A. A. and Tsygankov, S. S., Astronomy Letters, 35, 433 (2009)}\phantom{}{}

\reference{\it Lutovinov, A. A., Tsygankov, S. S., Postnov, K. A., et al., MNRAS, 466, 593 (2017)}\phantom{}{}

\reference{\it Malacaria, C., Jenke, P., Wilson-Hodge, C. A., and Roberts, O. J., The Astronomer’s Telegram, 12614, 1 (2019)}\phantom{}{}

\reference{\it Meegan, C., Lichti, G., Bhat, P. N., et al., The Astrophysical Journal, 702, 791 (2009)}\phantom{}{}

\reference{\it Molkov, S., Lutovinov, A., Tsygankov, S., Mereminskiy, I., and Mushtukov, A., The Astrophysical Journal, 883, L11 (2019)}\phantom{}{}

\reference{\it Mukerjee, K., Antia, H. M., and Katoch, T., The Astrophysical Journal, 897, 73 (2020)}\phantom{}{}

\reference{\it Mushtukov, A. A., Suleimanov, V. F., Tsygankov, S. S., and Portegies Zwart, S., Monthly Notices of the Royal Astronomical Society, 503, 5193 (2021)}\phantom{}{}

\reference{\it Reig, P., Negueruela, I., Papamastorakis, G., Manousakis, A., and Kougentakis, T., Astronomy \& Astrophysics, 440, 637 (2005)}\phantom{}{}

\reference{\it Sokolova-Lapa, E., Gornostaev, M., Wilms, J., et al., Astronomy \& Astrophysics, 651, A12 (2021)}\phantom{}{}

\reference{\it Titarchuk, L., AIP Conf. Proc., No. 304, 380 (1994)}\phantom{}{}

\reference{\it Wilson, C. A., Finger, M. H., Harmon, B. A., Chakrabarty, D., and Strohmayer, T., The Astrophysical Journal, 499, 820 (1998)}\phantom{}{}

\reference{\it Wilson, C. A., Weisskopf, M. C., Finger, M. H., et al., The Astrophysical Journal, 622, 1024 (2005)}\phantom{}{}

\reference{\it Harrison, F. A., Craig, W. W., Christensen, F. E., et al., The Astrophysical Journal, 770, 103 (2013)}\phantom{}{}

\reference{\it Tsygankov, S. S., Doroshenko, V., Mushtukov, A. A., et al., Monthly Notices of the Royal Astronomical Society: Letters, 487, L30 (2019a)}\phantom{}{}

\reference{\it Tsygankov, S. S., Lutovinov, A. A., Churazov, E. M., and Sunyaev, R. A., Astronomy Letters, 33, 368 (2007)}\phantom{}{}

\reference{\it Tsygankov, S. S., Lutovinov, A. A., and Serber, A. V., Monthly Notices of the Royal Astronomical Society, 401, 1628 (2010)}\phantom{}{}

\reference{\it Tsygankov, S. S., Wijnands, R., Lutovinov, A. A., et al., Monthly Notices of the Royal Astronomical Society, 470, 126 (2017)}\phantom{}{}

\reference{\it Tsygankov, S. S., Rouco Escorial, A., Suleimanov, V. F., et al.,  Monthly Notices of the Royal Astronomical Society: Letters, 483, L144 (2019b)}\phantom{}{}

\reference{\it Evans, P. A., Beardmore, A. P., Page, K. L., et al., MNRAS, 397, 1177 (2009)}\phantom{}{}

\end{references}




\clearpage
\clearpage


\end{document}